\documentclass[
reprint,
amsmath,amssymb,
aps, prapplied,
superscriptaddress,
longbibliography,
floatfix,
]{revtex4-2}

\usepackage{graphicx} % Required for inserting images
\usepackage{dcolumn}
\usepackage{bm}
\usepackage[hidelinks]{hyperref}
\usepackage[utf8]{inputenc}
\usepackage[T1]{fontenc}
\usepackage{placeins}

\begin{document}
\preprint{APS/123-QED}
\title{Characterization and benchmarking of a phase-sensitive two-qubit gate using direct digital synthesis}

\author{Mats O. Tholén}
\affiliation{Nanostructure Physics, KTH Royal Institute of Technology, 106 91 Stockholm, Sweden}
\affiliation{Intermodulation Products AB, 823 93 Segersta, Sweden}

\author{Riccardo Borgani}
\affiliation{Intermodulation Products AB, 823 93 Segersta, Sweden}

\author{Christian Kri\v{z}an}
\author{Jonas Bylander}
\affiliation{Department of Microtechnology and Nanoscience, Chalmers University of Technology, 412 96 Gothenburg, Sweden}

\author{David B. Haviland}
\affiliation{Nanostructure Physics, KTH Royal Institute of Technology, 106 91 Stockholm, Sweden}
\email[The author to whom correspondence may be addressed: ]{haviland@kth.se}

\begin{abstract}
We implement an iSWAP gate with two transmon qubits using a flux-tunable coupler. Precise control of the relative phase of the qubit-control pulses and the parametric-coupler drive is achieved with a multi-channel instrument called Presto using direct digital synthesis (DDS), a promising technique for scaling up quantum systems. We describe the process of tuning and benchmarking the iSWAP gate, where the relative phase of the pulses is controlled via software. We perform the iSWAP gate in 290 ns, validate it with quantum-state tomography, and measure 2\% error with interleaved randomized benchmarking.
\end{abstract}

\maketitle

\section{Introduction}
In 1982 Richard Feynman proposed using quantum-mechanical phenomena to perform calculations that would be impractical or impossible using classical computers \cite{FeynmanRichardP.1982Spwc}.
The idea to build a programmable quantum computer and use it for computation has evolved into several branches, one being superconducting electrical circuits at low temperature acting as tunable artificial atoms. The effort to create a universal quantum computer accelerated significantly in 1994 when Peter Shor proved that a quantum computer could factor integers in polynomial time \cite{ShorP.W.1994Afqc}. Today quantum computing is developing rapidly, while being limited by the noisy intermediate-scale quantum (NISQ) hardware used for its physical implementation \cite{PreskillJohn2018QCit}.

General-purpose quantum computing requires a universal set of quantum gates \cite{alma99526009102456}. The state of a qubit is represented as a point on the Bloch sphere, and any single-qubit gate is realized by a combination of resonant pulses that drive rotations by $\pi/2$ around the X axis of the sphere, and of virtual Z rotations implemented by changing the reference phase of all subsequent pulses \cite{McKayDavidC.2017EZgf}, effectively producing rotation around the Z axis. In addition to single-qubit gates, universality requires at least one two-qubit entangling gate, for example the CZ or iSWAP gates implemented with superconducting qubits \cite{GanzhornM.2019GSoM,PhysRevA.96.062323, PhysRevB.79.020507, KrantzP2019Aqeg}. The CZ gate does not require control of the relative phase of the resonant pulses, and is therefore potentially simpler to implement, but some algorithms benefit in terms of reduced circuit depth (number of sequential pulses) by using the iSWAP gate which does require phase control \cite{PhysRevA.98.022322, otterbach2017unsupervised}. Such benefit is especially important in the NISQ-era of quantum computing, where long pulse sequences degrade the achievable fidelity.

Contrary to rotating single qubits from a known state, where the phase of the first control pulse can be arbitrary, the iSWAP gate implemented with a resonant coupler requires control of the relative phase of three pulses: two control pulses for each single-qubit rotation and the pulse to the coupler. 
Generating these control pulses at microwave frequencies is usually a two-step process where pulse envelopes are synthesized at an intermediate frequency and each mixed with a dedicated local oscillator (LO) \cite{Heinsoo2018}. 
The phase of the resulting pulse depends on the phase of the envelope and the phase of the LOs whose frequencies must be locked with a deterministic phase relation. Pulses synthesized with general-purpose arbitrary-waveform generators and independent microwave LOs require additional synchronization hardware to fulfill these requirements \cite{ganzhorn2020benchmarking}. 

Recent developments in high-speed digital electronics lead to the Zynq UltraScale+ RFSoC by AMD (formerly Xilinx) \cite{rfsoc}, a chip targeted at software-defined radio for the 5G telecom development. The RFSoC family of chips contains multiple digital-to-analog and analog-to-digital converters with high enough sample rate and bandwidth to directly synthesize and measure microwave signals in the 4-8 GHz band typically used for superconducting qubits. The RFSoC has quickly gained popularity in the field of quantum computing, resulting in projects such as QICK \cite{stefanazzi_qick_2022} and QiCells \cite{GebauerRichard2023QAMR}. In this work we characterize and benchmark the iSWAP gate using a multi-channel digital microwave platform called Presto \cite{TholénMatsO.2022Maco}. 

Presto is based on the RFSoC and it allows for phase and frequency control across 16 output ports for arbitrary pulse sequences controlled via a Python application programming interface. With 16 channels synthesizing signals directly in the 4-8 GHz band, Presto replaces 32 AWG channels and 16 LOs driving 16 IQ mixers. Direct digital synthesis furthermore avoids mixer calibration, LO leakage and other distortion products. Presto also has 16 inputs which directly sample in the 4 - 8 GHz band, replacing down-converting LOs and IQ mixers. DDS thus provides several benefits when scaling to a larger number of qubits \cite{raftery2017direct}.

\section{The iSWAP gate}

\begin{figure}
  \centering
  \includegraphics{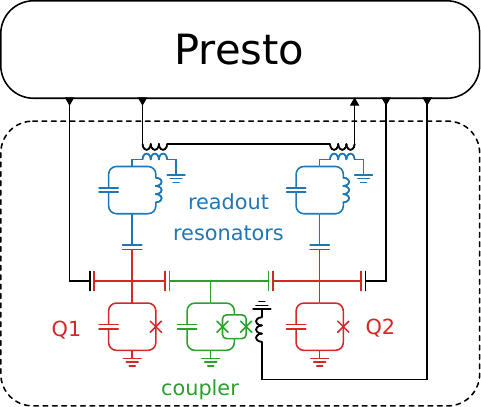}
  \caption{The sample used contains two transmon qubits (red) each with a control line and a readout resonator (blue). The readout resonators are inductively coupled to a shared control line for multiplexed readout. The qubits are capacitively coupled to a parametric coupler (green) whose frequency is DC biased and AC modulated using a separate flux line. All features needed to control and readout the qubits, synchronize clocks and trigger the pulse sequence are provided by Presto, making the hardware setup relatively simple.
  }
  \label{fig:setup}
\end{figure}

The iSWAP gate is a symmetric two-qubit gate that exchanges the two qubit states, $\left|\mathrm{ab}\right> \rightleftarrows \left|\mathrm{ba}\right>$, and multiplies the amplitudes of $\left|01\right>$ and $\left|10\right>$ by $i$. The matrix representation of the gate is
\begin{center}
$iSWAP = \begin{pmatrix}
1 & 0 & 0 & 0\\
0 & 0 & i & 0\\
0 & i & 0 & 0\\
0 & 0 & 0 & 1
\end{pmatrix}$.
\end{center}

We implement the iSWAP gate with a superconducting circuit schematically shown in Fig.~\ref{fig:setup}.  The circuit has two fixed-frequency transmon qubits Q1 and Q2, with individual control lines and readout resonators.
Both qubits are capacitively coupled to a common resonator through which the quantum states are swapped. The coupler is  both DC and AC modulated with an on-chip flux line.

The iSWAP gate is realized by modulating the coupler at the difference frequency of the two qubits $\omega_c = \omega_{Q1}-\omega_{Q2}$, driving the $\left|\mathrm{01}\right> \rightleftarrows \left|\mathrm{10}\right>$ transition at a rate $\Omega$ determined by the modulation amplitude \cite{PhysRevA.96.062323,ganzhorn2020benchmarking}.  The resulting unitary state transformation is
\begin{center}
\begin{equation}
\label{equ:unitary}
U(\Omega\tau, \eta) = \begin{pmatrix}
1 & 0 & 0 & 0\\
0 & \cos\frac{\Omega\tau}{2} & ie^{i\eta}\sin\frac{\Omega\tau}{2} & 0\\
0 & ie^{-i\eta}\sin\frac{\Omega\tau}{2} & \cos\frac{\Omega\tau}{2} & 0\\
0 & 0 & 0 & 1
\end{pmatrix},
\end{equation}
\end{center}
where $\tau$  is the duration of the modulation pulse, and $\eta$ is the phase difference between the coupler drive at $\omega_c$ and the difference of the two qubits' phases. 

The iSWAP is realized when $\Omega\tau = \pi$ and $\eta = 0$. Calibration of the iSWAP gate thus requires finding a suitable coupler drive amplitude, duration and phase to fulfill these conditions.

\section{Controlling microwave phase}

\begin{figure}
  \centering
  \includegraphics{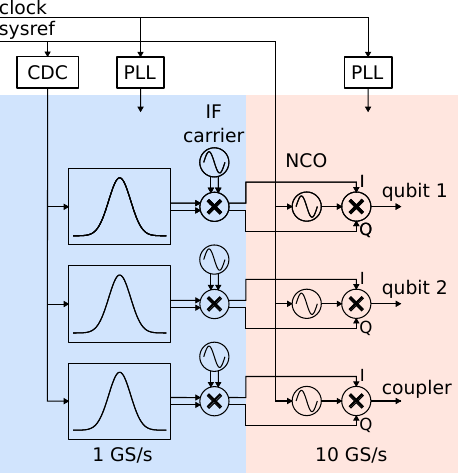}
  \caption{
  The infrastructure provided by the RFSoC enables synchronizing the NCOs and thereby controlling the relative phase of the microwave signals synthesized at different ports.  The RFSoC also provides infrastructure to transfer and interpolate data between the 1 GS/s clock domain to the 10 GS/s clock domain. In the 1 GS/s clock domain, the FPGA runs at a 500 MHz clock frequency and generates two complex-valued samples every clock cycle. All signals are generated synchronously and signal generation is started by the same sysref signal that synchronizes the NCOs, after a clock domain crossing (CDC). All clocks used are generated from the same base clock. The phase of the intermediate frequency (IF) carrier generators can be changed at any 2 ns interval.
  }
  \label{fig:generator}
\end{figure}

Phase-sensitive gates require precise control of the relative phase of signals on three separate channels: the two control lines to the qubits, and the flux line to the coupler.
Control of the relative microwave phase of multiple signals is comprised of two requirements: phase-locked microwave sources, and the ability to start all channels in a gate sequence with a known, or at least repeatable, phase relation between all microwave sources. When using separate instruments for the three control signals, the instruments have to be synchronized in frequency and phase, and the pulse sequence must to be started (triggered) at a known state of the relative phase of all signals \cite{ganzhorn2020benchmarking}.
This rigid phase control is not automatically obtained with the RFSoC, but the RFSoC provides a platform for its realization through software and firmware, a technology commonly referred to as 'software defined radio'. 
Since Presto derives all signals from one master clock on the RFSoC, and also synthesizes signals directly in the microwave band without analog upconversion, it readily achieves synchronization of all ports in frequency and phase.

As shown in Fig.~\ref{fig:generator}, Presto's field-programmable gate array (FPGA) runs at a 500 MHz clock frequency and generates signals at an intermediate sampling rate of 1 GS/s (two complex-valued samples per channel at each clock cycle). The signals are then interpolated to the sampling rate of the digital-to-analog converter and digitally upconverted by multiplying with a numerically-controlled oscillator (NCO) to reach the target frequency. Rigid phase control requires aligning the phases of the NCOs while synchronously starting the generation of the intermediate-frequency (IF) signals.

The first requirement for controlling microwave phases, phase locking the RF sources, is handled by a procedure known as multi-tile synchronization (MTS) which is a built-in feature of the RFSoC.
MTS is a simplified version of the JEDEC standard JESD204B (for data-converter synchronization and communication) implemented by Xilinx.  Presto uses MTS to synchronize all channels within an instrument through distribution of a common clock and a phase reference called `sysref'.  Synchronization of multiple Presto units is achieved through distribution of a clock-sysref pair with a separate instrument called Metronomo.  Further details can be found in Ref.~\cite{TholénMatsO.2022Maco}. 

The second requirement to microwave phase control, starting the IF signal generators with a known phase relation, is realized by triggering the start of the IF signal generation with sysref. This synchronization is simplified by the relatively low sampling rate of the IF signals (500 MS/s), but it requires properly-aligned synchronous transfer of the sysref signal to the IF clock domain. Once the NCOs are synchronized and the sequence is started, the phase of each output is digitally controlled by the IF signal.  It is possible to update both phase and frequency every 2 ns as these are defined in the 500 MHz IF clock domain.

Both NCO synchronization and starting of a measurement sequence are achieved with single sysref pulses triggered by software, not by the real-time signal-generation system.
When a measurement sequence needs to be run repeatedly with the same initial phase relation, for example when averaging, synchronization using sysref adds a significant amount of dead time to each repetition.  Synchronization in the IF domain comes with much lower dead time, requiring 2 ns as determined by the 500 MHz FPGA clock.  Therefore, a better strategy for repeated measurements is to split the frequencies of all the generated pulses into an NCO part and a remainder which is synthesized in the IF domain. The NCO frequency is fixed at an exact integer multiple of the repetition rate, so that its phase will start at the exact same value in every instance of the measurement sequence.  All frequency and phase changes in the sequence are done in the IF domain where synchronization requires only 2 ns.

\section{Measurements}
\label{sec:meas}

\begin{figure*}
  \centering
  \includegraphics{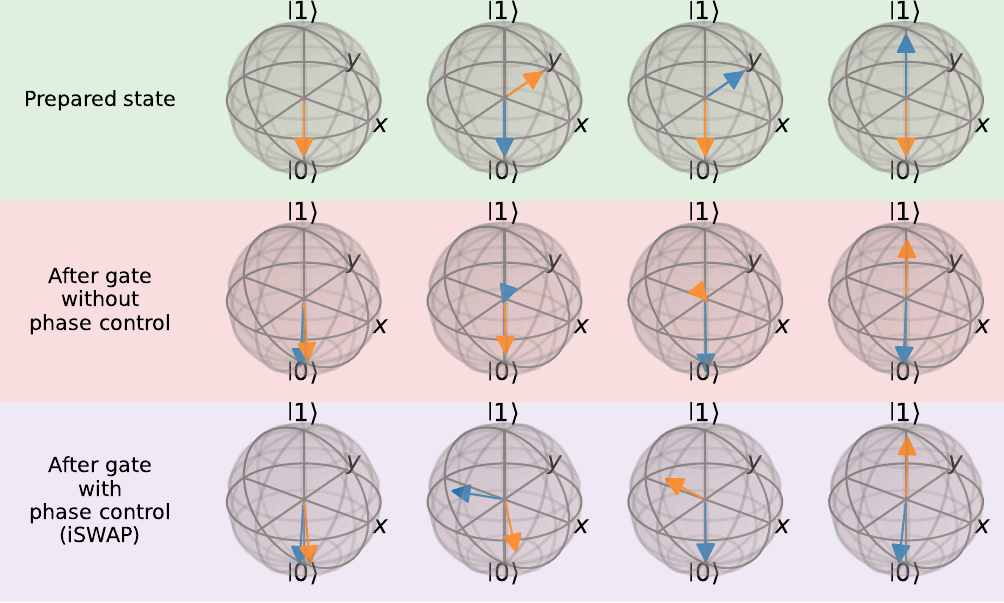}
  \caption{
    Tuning the phase of the coupler drive. Top row shows the prepared state, with orange and blue arrows pointing to the state of qubits Q1 and Q2, respectively. Middle row the shows the measured state after performing the gate without proper control of the phase of the coupler drive: the phase acquired during the gate is different each time and after averaging the states sensitive to phase average to zero. Bottom row shows the measured state with proper control of the phase of the coupler drive.
  }
  \label{fig:bloch}
\end{figure*}

\begin{figure}
  \centering
  \includegraphics{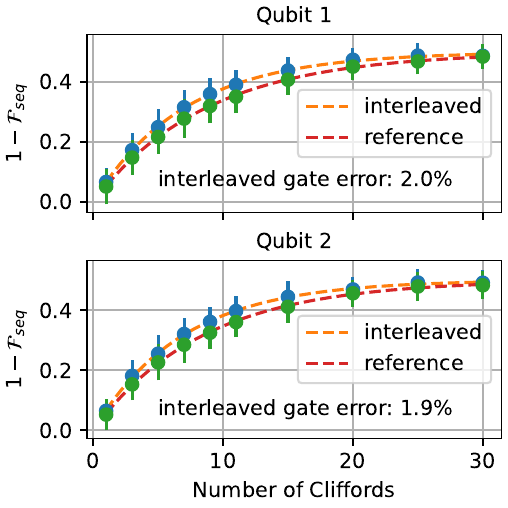}
  \caption{
    Interleaved randomized benchmarking of the iSWAP gate using 10 different sequence lengths and 200 circuit realizations for each sequence length. The two qubits are measured simultaneously at the end of each sequence with frequency-multiplexed readout. Blue and green circles depict the interleaved and reference sequence, respectively. Dashed lines are best fit.
  }
  \label{fig:irb}
\end{figure}

The initial procedure for calibrating the pulse parameters for the single-qubit gates and coupling pulses for two-qubit gates is described in detail in Appendix~\ref{apx:char}.  When the qubits are initialized to $\left|01\right>$ or $\left|10\right>$, the phase of the coupler drive does not affect the outcome of the two-qubit gate.  We therefore use these two initial states to adjust the amplitude and duration of the coupler drive needed to perform the swap operation.
However,  when either of the qubits is initialized to a superposition state, the phase of the final state does depend on the phase of the coupler drive. This phase dependence is clearly shown by the middle row of Fig.~\ref{fig:bloch} where the final state, having a different phase with each instance of the gate, averages to the center of the Bloch sphere if the phase of the coupler drive is not reproducible.

Tuning the duration of the coupler drive, we find that the optimal drive frequency is offset from the difference frequency of the two qubits. The presence of the coupler drive induces an effective shift of the center frequency of the coupler, in turn dispersively shifting the frequency of the qubits and causing them to accumulate a phase error during the gate. The accumulated error can be measured by applying two gates: one iSWAP and one -iSWAP. The second gate reverses the effect of the first gate, but the final state differs from the initial state by a phase error due to the dispersive shift \cite{ganzhorn2020benchmarking}. We compensate for this phase error by applying a virtual Z gate on both qubits at the end of every coupler drive.

Once the amplitude and duration (290 ns) of the coupler pulse are calibrated, we proceed to tuning the coupler phase $\eta$.  We prepare two initial superposition states, $\left|00\right> + i\left|01\right>$ and $i\left|10\right> + \left|00\right>$, apply the coupler pulse, and perform state tomography of the two qubits. We repeat this procedure while stepping the phase of the coupler drive until the gate has the expected outcome of an iSWAP gate.

Having tuned the gate parameters and verified the gate with simple circuits, we perform interleaved randomized benchmarking to estimate the gate fidelity. We use Qiskit \cite{Qiskit} to generate 200 circuit realizations for 10 different circuit depths from 1 to 30 Clifford gates.  We run each realization 1000 times, averaging the measured population of the qubits at the end of each sequence. For a 290-ns iSWAP gate and 20-ns single-qubit gates we fit a gate error of 2.0\% on one of the qubits and 1.9\% on the other, see Fig.~\ref{fig:irb}.

\section{Conclusions}
We demonstrated the use of direct digital synthesis (DDS) to implement an iSWAP gate.  After determining circuit parameters we implemented interleaved randomized benchmarking, achieving 2\% gate error. Direct digital synthesis in the microwave band simplifies hardware control of the gate by eliminating complicated analog methods of frequency up-conversion and trigger-based synchronization schemes, replacing them with software programming.  We demonstrated this with an instrument called Presto based on the RFSoC chip from AMD. Connecting Presto directly to cryostat cables fitted only with filters and amplifiers, we performed experiments where the relative phase of multiple signals was rigidly controlled through Python programming. An additional benefit of the tight integration of DDS-based signal generation is the ability to control 16 microwave ports using a single 2U rack mount instrument, decreasing system size and complexity, as well as the cost of controlling multi-qubit quantum processors.

\begin{acknowledgments}
This work was funded by the Knut and Alice Wallenberg Foundation (KAW) through the Wallenberg Centre for Quantum Technology (WACQT).  We thank Andreas Bengtsson for fabricating the qubit sample.  The authors
MOT, RB and DBH are part owners of Intermodulation Products AB, which manufactures and sells the Presto microwave platform used in this manuscript.
\end{acknowledgments}

\appendix
\section{Gate calibration}
\label{apx:char}

\begin{figure}
  \centering
  \includegraphics{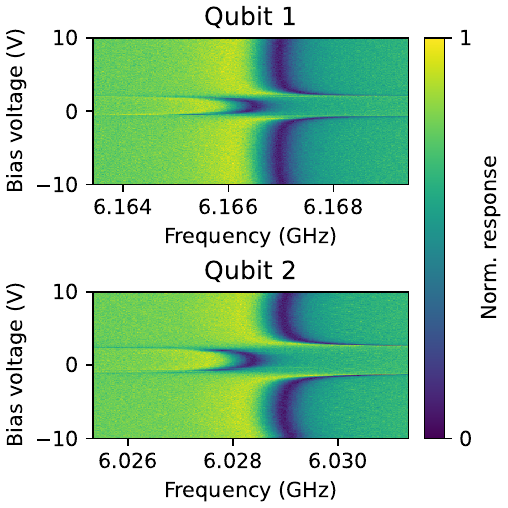}
  \caption{
  Spectroscopy of the readout resonators for different DC-bias voltages on the coupler. The avoided level crossings indicate when the coupler frequency coincides with the frequency of the readout resonators.  Measured response is normalized to the minimum and maximum values.
  }
  \label{fig:resonator_sweep}
\end{figure}

\begin{figure}
  \centering
  \includegraphics{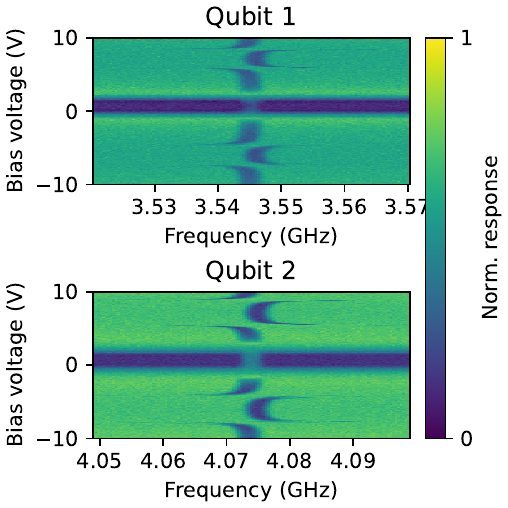}
  \caption{
  Two-tone qubit spectroscopy for different DC-bias voltages on the coupler. The avoided level crossings indicate when the coupler frequency coincides with the frequency of the qubits.  Measured response is normalized to the minimum and maximum values.
  }
  \label{fig:two_tone}
\end{figure}

\begin{figure}
  \centering
  \includegraphics{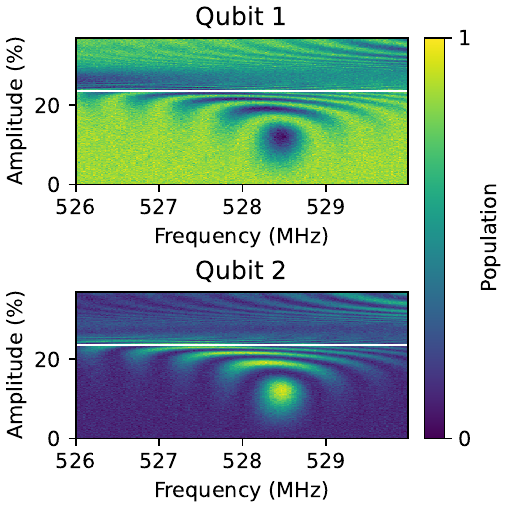}
  \caption{
  Population on the two qubits as function of coupler drive amplitude and frequency. White line indicates the selected amplitude, 23.6\% of the DAC full-scale range.
  }
  \label{fig:basweep}
\end{figure}

\begin{figure}
  \centering
  \includegraphics{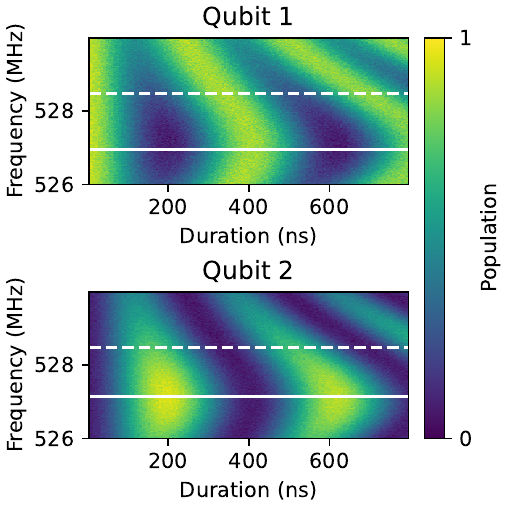}
  \caption{
    Tuning the duration of the coupler drive. The solid white line marks the coupler frequency that generates maximum contrast in population swapping. Due to dispersive shifts, this frequency is offset from the difference frequency of the two undriven qubits (dashed white line). The first maximum in population swap marks the duration of the iSWAP gate.
  }
  \label{fig:swap}
\end{figure}

Calibration of the single-qubit gates is described in Ref.~\citenum{TholénMatsO.2022Maco}. To find the optimal operating point of the flux-tunable coupler, we perform resonator spectroscopy (Fig.~\ref{fig:resonator_sweep}) and two-tone qubit spectroscopy (Fig.~\ref{fig:two_tone}), both while sweeping the DC flux applied to the coupler.
The avoided level crossings in the former measurement identify the DC-flux value at which the coupler resonance frequency coincides with that of the readout resonators. In the latter measurements, the avoided level crossings occur when the coupler resonance coincides with that of the qubits. We choose a DC-bias operating point of 3.775 V corresponding to a flux bias of 0.25 $\Phi_0$, the operating point which maximizes the difference in frequency between the coupler resonance and the resonances of the readout resonators and the qubits.

Calibrating the iSWAP gate requires finding the amplitude $\Omega$, duration $\tau$ and phase $\eta$ of the AC drive to the coupler. We perform this calibration with three successive measurements.
To calibrate $\Omega$, we initialize the qubits to $\left|\mathrm{10}\right>$, apply an AC drive to the coupler for a fixed duration of $\tau=2$~ \textmu s and measure the population of the qubits.   We repeat this measurement while sweeping the amplitude and frequency of the coupler drive (see Fig.~\ref{fig:basweep}).
From Eq.~(\ref{equ:unitary}), the resulting population on qubit Q1 and Q2 is $\cos^2 \frac{\Omega \tau}{2}$ and $\sin^2 \frac{\Omega \tau}{2}$, respectively, independent of the phase $\eta$.
Figure~\ref{fig:basweep} shows how the population swaps between the two qubits, and oscillates with increasing amplitude of the coupler pulse. For large amplitudes, the oscillations become too sensitive to amplitude and insensitive to frequency. We select an amplitude of $\Omega = 23.6\%$ of the DAC range, within the region where oscillations in the qubit population are well visible.

To calibrate $\tau$, we repeat the previous experiment with fixed amplitude $\Omega$ and with varying frequency and duration of the coupler drive as shown in Fig.~\ref{fig:swap}. From this plot we choose the frequency $\omega_C=527.05$~MHz with maximum contrast and the duration $\tau = 290$~ns with highest population exchange. Finally, we calibrate $\eta$ as described in the main text in Sec.~\ref{sec:meas}.

As visible in Fig.~\ref{fig:swap}, the optimal drive frequency $\omega_C$ is offset from the difference of the qubits' frequencies $\omega_{Q1} - \omega_{Q2}$. This shift is due to the dispersive interaction between the qubits and the coupler during the iSWAP gate. We measure the phase error caused by the dispersive shift by performing the gate twice, the second time with the opposite sign of the coupler pulse (-iSWAP). This sequence cancels the population swap, but doubles the phase shift acquired by the qubits.  We then compensate for this phase error by applying a virtual $Z$ gate after each iSWAP when performing randomized benchmarking.

\bibliography{mybib}

\end{document}